\documentclass[12pt]{article}
\usepackage{zbarskystyle}
\usepackage{stmaryrd}
\usepackage{bbm}
\usepackage{graphicx}
\title{Asymptotically faster algorithm for counting self-avoiding walks and self-avoiding polygons}
\author{Samuel Zbarsky, Princeton University}
\begin{document}
\maketitle
\begin{abstract}
We give an algorithm for counting  self-avoiding walks or self-avoiding polygons of length $n$ that runs in time $\exp(C\sqrt{n\log n})$ on 2-dimensional lattices and time $\exp(C_dn^{(d-1)/d}\log n)$ on $d$-dimensional lattices for $d>2$.
\end{abstract}
\section{Introduction}
Given a lattice, a self-avoiding walk (SAW) of length $n$ is a walk in the graph that starts at the origin and never revisits a vertex. A  self-avoiding polygon (SAP) is a SAW except that it returns to the origin on the $n$th step (also, when talking about counting SAPs, we do not care about which point on the polygon is the origin). SAWs and SAPs are used as a model for polymers in statistical mechanics. A good source for information on SAPs and SAWs (current as of 1993) is \cite{MadrasSlade}.

It is strongly believed (but not proved) that the number of SAWs of length $n$ on a lattice is given by $c_n\sim A n^{\gamma-1} \mu^n$ and the  number of SAWs of length $n$ is given by $p_n\sim B n^{\alpha-3} \mu^n$ where $\gamma$ and $\alpha$ are critical exponents depending only on dimension, while $\mu$ depends on the lattice, with $\mu\approx 2.6$ for the square lattice and $\mu\approx 4.7$ for the simple cubic lattice.

The naive algorithm for counting SAWs and SAPs is based on backtracking to list each individual path, so it has exponential complexity $\mu^n$. However, there have been a series of improved algorithms that run in time $C^n$ for various improved values of $C$ (omitting polynomial factors in the runtime). We will mostly just talk about the work for square and cubic lattices, since more work has been done on them and they are easier to work with, although there has been work on other lattices also. Most known algorithms can be adapted to any lattice of the same dimension. In $\cite{Enting80}$, Enting first applied the transfer matrix method to this problem for the square lattice (we will describe the method more in the next section). Variants of this method have been developed. Most recently, Clisby and Jensen \cite{ClisbyJensen12} counted all SAPs on the square lattice up to length 130 with an algorithm that had $C\le 1.2$ (it improves over a previous algorithm \cite{GuttmannJensen99} which had $C\approx 1.2$, but they do not specify what value of $C$ they have). Jensen \cite{Jensen13} counted all SAWs of length up to 79 on the square lattice with an algorithm that had $C\le1.334$ (it improves over a previous algorithm \cite{Jensen04} which had $C\approx 1.334$). There was also an algorithm for counting SAWs by Conway, Enting, and Guttman \cite{ConwayEntingGuttman93} that had $C=3^{1/4}\approx 1.316$, but it was slower in practice than Jensen's algorithm.

On three dimensional lattices, the best previously known algorithm for SAWs is the doubling method of Schram, Barkina, and Bisseling~\cite{Schrametal11, Schrametal17}. This method involves listing all self-avoiding walks of length $n/2$ and then using inclusion-exclusion to count nonintersecting pairs of such paths (corresponding to walks of length $n$). They have $C=\sqrt{2\mu}$ on any lattice, which for the cubic lattice gives $C\approx 3.06$ and allows them to count SAWs of length up to 36. The best count of SAPs for the cubic lattice is in \cite{ClisbyLiangSlade07}, which goes up to length 32.

In Section~\ref{sec:2d}, we give an algorithm for enumerating  SAPs for all lengths up to $n$ on the square lattice that runs in $\exp(C\sqrt{n\log n})$ time. In Section~\ref{sec:>2d},  we explain how to modify the algorithm to count SAPs for all lengths up to $n$ on the  lattice $\Z^d$ so that it runs in time $\exp(C_dn^{(d-1)/d}\log n)$. In Section~\ref{sec:practicality}, we give a few remarks on ideas for optimizing these algorithms for the square lattice for practical use.

This is a significant theoretical improvement over existing exponential-time algorithms. The algorithms presented here can, with minor modification, work for counting SAWs, as well as counting SAWs and SAPs with various statistics, or in certain convex regions in space, or on different $d$-dimensional lattices. The basic idea can probably also be used for some other transfer matrix enumeration problems, as long as the objects being counted are sparse in a suitable sense---in the case of SAPs, the sparsity lies in including only $n$ edges out of $\approx n^2$ potential edges in the rectangle.

\section{Better algorithm for two dimensions}\label{sec:2d}
We present the following algorithm for counting self-avoiding polygons on $\Z^2$ in time\\
$\exp(C\sqrt{n\log n})$. With slight modifications that may affect the constant $C$, but not the form of the asymptotics, we can use the same algorithms to count self-avoiding walks, work on other lattices, count SAWs or SAPs restricted to some convex or relatively nice region of space, or count SAWs or SAPs satisfying certain statistics (such as mean square end-to-end distance for SAPs, or the sum of coordinates of vertices and sum of their squares, which together allow you to calculate the mean square distance between pairs of point vertices). Below, we will give a brief description of the transfer matrix method as applied to this problem, and then how we modify it. For a detailed description of how the transfer matrix algorithm works in the case of SAPs, see \cite{ClisbyJensen12,Enting80,EntingJensen09}.

We will make no attempt in this section to optimize the constant $C$ in the runtime. The algorithm works based on the transfer matrix method. We will count SAPs inscribed in an $L$ by $W$ rectangle and then sum over $L,W\le n$. At each step, we will have some set of vertices in a set $A$ so that if any vertex is in $A$, then the vertex to its left and above it are also in $A$. $A$ will start out with only vertices above and to the left of our rectangle, and we will add one vertex at a time to $A$. The state space will be the list of edges connecting $A$ to $\bar A$, how these edges are connected by paths inside $A$, how many edges of the SAP are contained inside $A$, and  whether the part of the SAP contained in $A$ touches each of the four edges of the rectangle. For instance, one state is given by the figure below along with the caption.
\begin{figure}[h!]
  \centering
\includegraphics[scale=0.3]{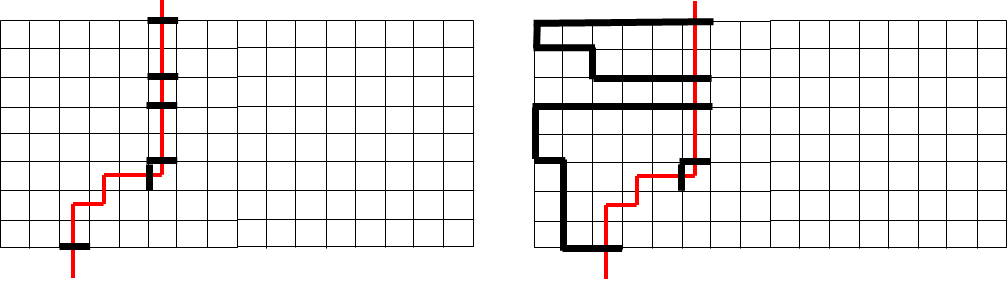}
\caption{The red line is the boundary of $A$. The black edges crossing it in the figure on the left are part of the state. The rest of the state is the fact that the path touches both the bottom and top of the rectangle, that there are 24 edges in the interior of $A$, and that among the edges shown, the first two connect, the 3rd connects to the 6th, and the 4th and 5th connect. The figure on the right is one possible arrangement inside $A$, so it contributes 1 to the number stored corresponding to this state.}
\end{figure}

If the number of states is $S$, then we will store a vector of $S$ numbers, namely how many arrangements inside $A$ achieve every one of the states. It is easy to come up with fast local rules for updating when a new vertex is added to $A$, and these rules are spelled out in detail in \cite{ClisbyJensen12} for counting SAPs and in \cite{Jensen13} for counting SAWs. These rules are equivalent to multiplying our state vector by some matrix, hence the name ``transfer matrix method". However, because all of the rules are local, this matrix is sparse, which speeds up the updating step. Previous algorithms \cite{ClisbyJensen12,Jensen13} to use the transfer matrix method updated $A$ by moving the boundary to the right one column at a time as follows (the red line shows the bottom right boundary of $A$):
\begin{figure}[h!]
  \centering
\includegraphics[scale=0.3]{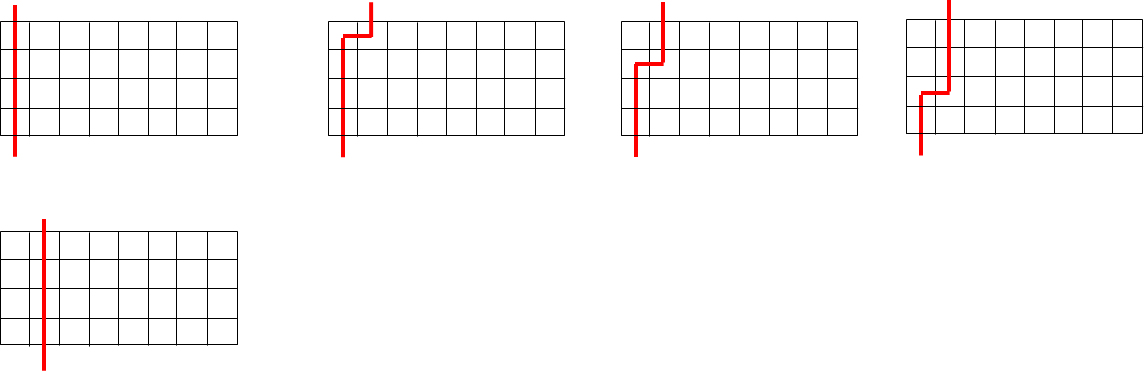}
\caption{Order of expanding $A$ in previous algorithms. The figures proceed left to right row by row.}
\end{figure}

Unfortunately, that led to them having exponential state space and thus exponential runtime (though better than the naive runtime).

We come up with a different way to move the boundary based on the following observation. Pick some $2\le k<n$ and let $q=\floor{n/k}$. For $P$ an SAP of length $n$ inscribed in our rectangle, let
\begin{align*}
C(P)=&\{\alpha\in \{0,\ldots,k-1\} \mid \text{every column whose number is $\alpha$ mod $k$ has at most $q$}\\
      &\qquad \qquad\qquad\text{ horizontal edges of the polygon crossing it}\}.
\end{align*}
Then by the pigeonhole principle, $C(P)$ is nonempty. We count the number of SAPs with nonempty  $C(P)$ (and thus the number of SAPs) using inclusion-exclusion. For each nonempty $K\subset \{0,\ldots,k-1\}$ it suffices to count the number of SAPs such that $K\subseteq C(P)$, and we will denote this count by $N_K$. We find $N_K$ by advancing the boundary of $A$ from column to column, skipping those columns whose number is $\beta$ mod $k$ for some $\beta\notin K$, and whenever we stop at a column that is $\alpha$ mod $k$ for $\alpha\in K$, we throw away those states that have more than $q$ edges crossing that column. For instance, if we skip one column, we do it as follows:
\begin{figure}[h!]
  \centering
\includegraphics[scale=0.3]{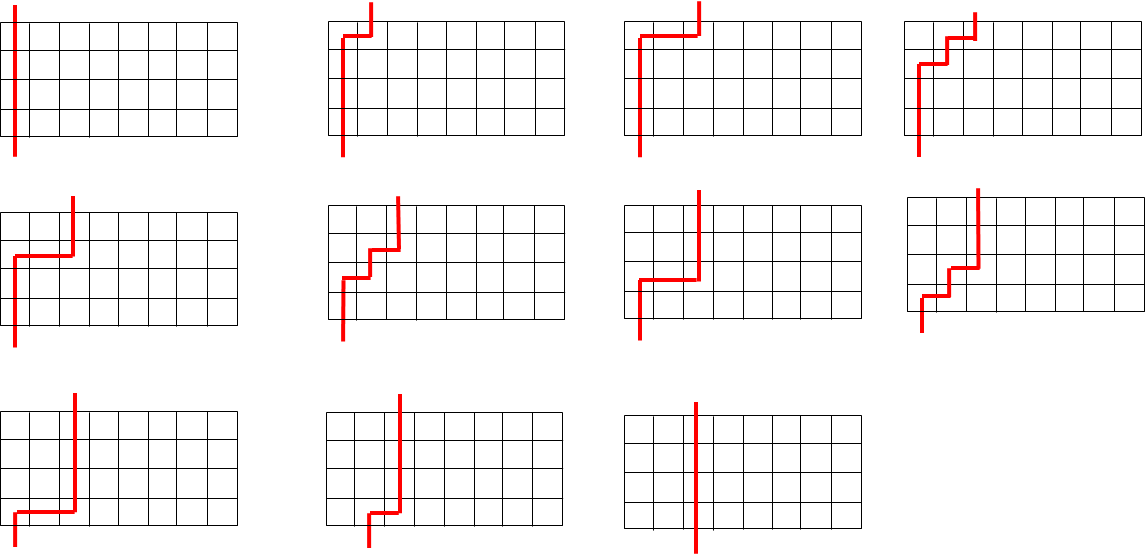}
\caption{Order of expanding $A$ in our algorithm when skipping a single column. The figures proceed left to right row by row.}
\end{figure}

For example, if $k=6$ and $K=\{1,2,4\}$, then we go through, stopping at the shaded columns:
\begin{figure}[h!]
  \centering
\includegraphics[scale=0.3]{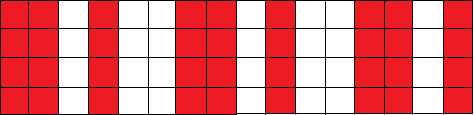}
\caption{Columns we stop at when advancing $A$ for $k=6$ when computing $N_{\{1,2,4\}}$}
\end{figure}

Then the total number of SAPs is given by the inclusion-exclusion formula
\[
c_n=\sum_{K\subseteq \{0,\ldots,k-1\}, K\ne \emptyset} (-1)^{|K|+1}N_K.
\]
For instance if $k=4$, then
\bal
N&=N_{\{0\}}+N_{\{1\}}+N_{\{2\}}+N_{\{3\}}-N_{\{0,1\}}-N_{\{0,2\}}-N_{\{0,3\}}-N_{\{1,2\}}-N_{\{1,3\}}-N_{\{2,3\}}\\
&\qquad+N_{\{0,1,2\}}+N_{\{0,1,3\}}+N_{\{0,2,3\}}+N_{\{1,2,3\}}-N_{\{0,1,2,3\}}
\eal
Note that when computing $N_K$, at each stage all but at most $k+1$ segments of the boundary of $A$ will lie in one of two columns that are $\alpha$ mod $k$ for $\alpha\in K$, so there are at most $2q+k+1$ edges crossing the boundary at any time and at most
\[
\binom{n}{q}^22^{k+1}
\]
possibilities for those edge locations. Furthermore, because the path cannot self-intersect, the loops which connect edges crossing the boundary cannot cross each other, so they form a balanced set of parentheses, so the number of ways of connecting on the left is bounded by Catalan numbers. Using a power of 2 as an upper bound for Catalan numbers, there are at most $2^{2q+k+1}$ ways these edges can be connected inside $A$. Thus the number of states is bounded by
\[
n^{O(1)}\binom{n}{q}^22^{k+1}2^{2q+k+1}\le 2^{O((n\log n)/k+k+\log n)}.
\]
Remembering that we need $O(n^2)$ update steps for each $K$ and there are $2^k-1$ values of $K$ we need to deal with, we get that the runtime is
\[
 2^{O((n\log n)/k+k+\log n)}
\]
and optimizing $k$ by setting it equal to $\sqrt{n\log n}$, we get a runtime of
\[
\exp(C\sqrt{n\log n}).
\]
\section{Better algorithm for more than two dimensions}\label{sec:>2d}
We use a similar algorithm for $d\ge 3$ with some modifications. First, instead of merely choosing the set of good columns $K$, we choose a set of good hyperplanes in each direction. Second, when advancing, instead of going in chunks of at most $k$ columns, we go in chunks where each chunk is a rectangular prism each of whose sides is at most $k$. Thus the size of the part of the boundary of $A$ not lying in good hyperplanes is at most $dk^{d-1}$, so we can bound the number of edges in each state by $s=2dn/k+dk^{d-1}$. Finally, we can no longer use non-self-intersection to limit how edges can be connected within $A$, so we get a factor corresponding to the number of matchings of $s$ elements, which is $(s-1)!!\approx 2^{Cs\log s}$ instead of $2^s$. We now optimize $k=n^{1/d}$ and get a runtime of
\[
\exp(C_dn^{(d-1)/d}\log n).
\]
Note that we are getting $\log n$ here instead of $\sqrt{\log n}$ as in the $d=2$ case because we are counting all possible matchings on $s$ elements. However, many of these will require more than $n$ edges to implement. Thus, if we prune as we go, keeping only those states that can potentially arise as part of a SAP, we will get far fewer states, potentially replacing the $\log n$ with $\log n^\sigma$ for some $\sigma<1$.

\section{Optimization considerations}\label{sec:practicality}
It is unlikely that the algorithm for $d=3$ will be faster than the naive algorithm for any $n$ for which the computation is currently feasible, since for those values of $n$, we have $n^{2/3}\log n>n$ and the algorithm presented here has a large polynomial factor in front. For $d>3$ it is even worse. However, there is a chance that the $d=2$ algorithm on the square lattice can improve on the record of $n=79$ for SAWs set by Jensen in $\cite{Jensen13}$. In this section we discuss some possible optimizations of the algorithm. First, one could speed the algorithm up by thinking of it as a modification of the algorithm given in that paper (and thus use all the optimizations used there, including pruning and keeping track of how things are connected on the right rather than the left). Additionally, by using that paper's framework of counting the number of paths inscribed in a rectangle of width $W$ with length $L\ge W$, one can get better bounds than $n/k$ for $q$ (since we need to use at least $2W$ edges to cross the rectangle from bottom to top and left to right). We can perhaps eke out a further improvement by keeping track of the number $B$ of vertical edges. If $B$ is close to $W$, we only have a few spare vertical edges to use as vertical edges crossing the boundary or to connect horizontal edges, which severely restricts the number of states $S$, so we will have large gains from pruning. If the $B$ is significantly larger than $W$, we are tying up $B+W$ edges in order to be inscribed in the rectangle, so we can use a smaller value of $q$. One can choose different values of $k$ and $q$ for different values of $B$ (or for ranges of values of $B$, to have to run the algorithm fewer times). It is not clear how to make these choices optimally. It is also not clear if these optimizations are enough to improve on the record of $n=79$.
\section{Acknowledgments}
The author would like to thank the anonymous referees for copious useful comments.
This material is based upon work supported by the National Science Foundation Graduate Research Fellowship Program under Grant No. DGE-1656466. Any opinions, findings, and conclusions or recommendations expressed in this material are those of the author and do not necessarily reflect the views of the National Science Foundation.
\bibliographystyle{abbrv}
\bibliography{zbarskybib}
\end{document}